\documentclass[aps,prl,twocolumn,superscriptaddress,showpacs]{revtex4}

\usepackage{graphicx}
\usepackage[usenames]{color}

\usepackage{amsmath}
\usepackage{amssymb}
\usepackage{bbm,ulem}

\newcommand{\br}{\mathbf{r}}
\newcommand{\bq}{\mathbf{q}}
\newcommand{\bk}{\mathbf{k}}

\newcommand{\calF}{\mathcal{F}}
\newcommand{\ells}{\ell_\text{s}}
\newcommand{\elltr}{\ell_\text{tr}}

\newcommand{\journal}[4]{#1 \textbf{#2}, #3 (#4)}
\newcommand{\PRL}[3]{\journal{Phys.\ Rev.\ Lett.}{#1}{#2}{#3}}

\newcommand{\bra}[1]{\left\langle #1\right|}
\newcommand{\ket}[1]{\left| #1\right\rangle}
\newcommand{\av}[1]{\overline{#1}}

\newcommand{\avV}{\av{V}}
\newcommand{\efield}{\ensuremath{\mathcal{E}}}

\newcommand{\power}{\ensuremath{\mathcal{P}}}

\newcommand{\fluct}{\ensuremath{\delta V}}

\renewcommand{\vec}[1]{\ensuremath{\mathbf{#1}}}

\newcommand{\refeq}[1]{(\ref{#1})}

\begin{document}

\title{Localization of Matter Waves in 2D-Disordered Optical Potentials}

\author{R. C. Kuhn}
\email[]{robert.kuhn@uni-bayreuth.de}
\affiliation{Physikalisches Institut, Universit\"at Bayreuth,
D-95440 Bayreuth}
\affiliation{Institut Non Lin\'eaire de Nice Sophia Antipolis, UMR
6618 du CNRS, 1361 route des Lucioles, F-06560 Valbonne}

\author{C. Miniatura}
\affiliation{Institut Non Lin\'eaire de Nice Sophia Antipolis, UMR
6618 du CNRS, 1361 route des Lucioles, F-06560 Valbonne}

\author{D. Delande}
\affiliation{Laboratoire Kastler Brossel, Universit\'e Pierre et
Marie Curie, 4 Place Jussieu, F-75005 Paris}

\author{O. Sigwarth}
\affiliation{Physikalisches Institut, Universit\"at Bayreuth,
D-95440 Bayreuth}

\author{C. A. M\"uller}
\affiliation{Physikalisches Institut, Universit\"at Bayreuth,
D-95440 Bayreuth}


\begin{abstract}
We consider ultracold atoms in 2D-disordered optical potentials and
calculate microscopic quantities characterizing matter wave quantum
transport in the non-interacting regime. We derive the diffusion
constant as function of all relevant microscopic parameters
and show that coherent multiple scattering induces significant weak
localization effects. In particular, we find that even the
strong localization regime is accessible with
current experimental techniques and calculate the corresponding localization
length.
\end{abstract}


\maketitle
\paragraph{Introduction:}

Ultracold atoms in optical potentials can be used to realize
condensed matter model systems in a very versatile manner
\cite{Jaksch05}. Having at hand the possibility to shape external
potentials almost at will, a natural direction of investigation is
the disorder driven superfluid-insulator transition
\cite{Fisher89} or the Anderson transition \cite{Kramer93}.
Interest in this transition has been
renewed
since the
experimental observation that even a small disorder in confining
fields leads to a fractioning of quasi-1D condensates in
waveguide structures on atom chips \cite{Schumm04}.
Very recently,
several studies of
Bose condensates in speckle potentials
\cite{Lye04}
have shown an efficient suppression of 1D transport by disorder
\cite{Clement05,Fort05,Schulte05}.

The physics of interacting particles in a disordered
environment 
has been discussed already for a number of years
for electrons and superfluid helium
\cite{Belitz94,Fisher89}. 
For cold atoms, recent contributions include a numerical
study of the Bose-Hubbard and Anderson hopping models
\cite{Damski03}, a
renormalization group approach in the case of off-diagonal
disorder \cite{Altman04}, a Bose-Fermi mapping for hardcore bosons
\cite{DeMartino05}, and a transfer matrix treatment of atomic
matter waves interacting with impurity atoms in an optical lattice
\cite{Gavish04}. All these approaches study 1D systems. Bose
condensates in 2D optical quasicrystal lattices were considered in
\cite{Sanchez-Palencia05}.
In this context, our aim is to concentrate on the effect of disorder
on quantum transport in the non-interacting regime, 
leaving the intriguing impact of interactions 
for later studies. 
Interaction effects are small in the low-density wings of expanding
Bose condensates \cite{Clement05} that are obtained from 
interaction-dominated condensates in the 
Thomas-Fermi regime by opening the trap potential. 
If one insists
on working in the high-density regime, 
single-particle dynamics can be studied by reducing
the scattering length via Feshbach resonances
\cite{Volz03}, eventually reaching the ideal Bose gas regime~\cite{Salomon02}.
Alternatively, one can work with spin-polarized fermions 
\cite{Fermions} whose collisions are blocked
by the Pauli principle. 

In this Letter, we report analytical results for the 2D dynamics of cold
atoms in a far-detuned optical speckle potential.
Using a perturbative Green's function approach, we calculate transport
quantities relevant for the diffusive regime and derive the
weak localization correction to the classical diffusion constant.
The 2D geometry is particularly advantageous for measuring quantum
corrections to classical transport because diffusive trajectories
always return to their starting point, thus favoring localization effects.
Making use of the microscopic characteristics of the speckle
potential instead of effective models of disorder,
we show that a highly disordered environment with strong
scattering can be tailored with speckle potentials. Furthermore, we find 
that the strong localization threshold can be reached with current
experimental techniques.

\paragraph{Intensity transport:}

A cloud of non-interacting cold atoms
is described
by the single particle Hamiltonian
$H=p^2/2m+V(\br)$ where $V(\br)$ is a static 2D random potential
after the harmonic confinement in the transport directions has been
switched off as realized in
\cite{Clement05,Fort05,Schulte05}. The initial state density
matrix $\varrho_0$ of the atomic cloud
evolves in time as $\varrho(t)= U(t) \,\varrho_0\,
U^\dagger(t)$ with the evolution operator
$U(t)=\exp(-iHt/\hbar)$. Meaningful
matter wave transport observables involve a statistical average over
all possible realizations of disorder. One important quantity is
the average probability density $p(\br,t)=\av{\bra{\br} \varrho(t)
\ket{\br}}$ of particles at point $\br$ and time $t$ (a bar
denotes the disorder average). Its Fourier transform
$p(\bq,\Omega)=\int d^2r
\int dt\, p(\br,t) \,\exp(i\Omega t-i\bq
\br)$ is given by
\begin{equation}
p(\bq,\Omega) = \int\frac{d^2k}{(2\pi)^2}\;
 \varrho_0(\bk,\bq) \,\Phi(\bk,\bq,\Omega),
\label{Pqeps.eq}
\end{equation}
where $\varrho_0(\bk,\bq)=
\bra{\bk+\bq/2}\varrho_0\ket{\bk-\bq/2}$ contains all information
about the initial atomic density distribution, and $\Phi$ is the
intensity relaxation kernel for plane waves with on-shell energy
$E=\hbar^2k^2/2m$. In the long time and
large distance limits $\Omega,q\to 0$, the 2D relaxation kernel
for isotropic intensity distributions has the characteristic pole
\begin{equation}
\Phi(\bk,\bq,\Omega) = \frac{1}{-i\Omega + D(k) q^2}
\end{equation}
that describes a diffusion process with diffusion constant
$D(k)$. Since diffusion solely relies on
the local conservation of particles and on linear response it
is a very robust phenomenon. In the
remainder of the paper, we will essentially calculate the
plane wave diffusion constant $D(k)$, including interference
corrections, as a function of all relevant microscopic parameters.
Once the diffusion constant is determined, the dynamics of any
initial density distribution is obtained by integrating
(\ref{Pqeps.eq}).

\paragraph{Speckle characteristics:}

Of particular importance are the characteristics of the optical
potential $V(\br) = \avV [1+ \fluct(\br)]$ with average $\avV$ and
normalized random component $\fluct(\br)$. The only
correlation function that we will need in the following is the
two-point correlator $ \power(\br) =
\av{\fluct(\br')\fluct(\br'+\br)}$.
We consider two-level atoms
(transition frequency $\omega_0$, transition width $\Gamma$,
saturation intensity $I_s$) exposed to a far-detuned monochromatic
speckle field $\efield(\br)$ at frequency $\omega_\text{L}=ck_\text{L}$ and
detuning $\delta=\omega_\text{L}-\omega_0$, generated from a laser
source with power $P$. The speckle pattern is created over a
surface of linear size $L$, the local field intensity being
$I(\br)=\epsilon_0c\,|\efield(\br)|^2/2$ with average value $I_\text{L}=P/L^2$.
The speckle optical dipolar potential is $V(\br) = \avV \,
I({\br})/I_\text{L}$, with $\avV=(\hbar\Gamma^2/8\delta)I_\text{L}/I_s$. This random
potential derives from the Gaussian random field
$\mathcal{E}(\br)$, but is not a Gaussian variable by itself. Its
pair correlator can be expressed as $\power(\br) =
|\gamma(\br)|^2$ with $\gamma (\vec{r})=\epsilon_0c\,\av{\efield^*(\br')
\efield(\br'+\br)}/2I_\text{L}$ being the normalized two-point field
correlation function \cite{goodman}. Its 2D Fourier transform
$\power(\bk)$ is the speckle power spectrum.

For a 2D speckle pattern produced by monochromatic illumination of
a holographic phase mask (transmission geometry) or of a rough
surface (reflection geometry), the far-field correlation reads
$\gamma(\vec{r}) = 2 J_1(u)/u$ where $u=r/\zeta$ and $J_1$ is the
first order Bessel function \cite{bornwolf}.
Here $\zeta=1/\alpha
k_\text{L}$ is the correlation length of the speckle potential,
and $\alpha=R/z \ll 1$ is the speckle aperture angle at a distance
$z$ from the speckle source with radius $R$.
The correlation length $\zeta$ defines the
intrinsic physical length scale of our system. In turn, it also
defines an energy scale for the atomic dynamics, $E_\zeta =
\hbar^2/m\zeta^2 = 2 \alpha^2 E_R $ in terms of the more familiar
recoil energy $E_R = \hbar^2k_\text{L}^2/2m$. The ratio $\eta
=\avV/E_\zeta$ measures the strength of the potential fluctuations
relative to the correlation energy $E_\zeta$. The 2D speckle power
spectrum is obtained as the convolution of two identical disks,
$\power(\bk) = 8 \calF(k\zeta/2)$ with $\calF(x)= \left[\arccos x -x
\sqrt{1-x^2}\right] \Theta(1-x)$. The Heaviside distribution
$\Theta$ reflects the fact that the potential is smooth on length
scales smaller than $\zeta$, but uncorrelated on distances larger
than $\zeta$.

\paragraph{Weak scattering regime:}

Microscopic transport parameters can be calculated using standard
diagrammatic Green's function techniques \cite{AkkerMon,diagrammatics,altshuler}. A
well controlled perturbative expansion is obtained if
the atomic energy $E$ lies above the mobility edge $E_c$:
\begin{equation}
\Delta= \frac{E_c}{E} < 1; \quad E_c =\frac{\avV^2}{
E_\zeta}=\eta^2 E_\zeta.\label{delta.eq}
\end{equation}
In physical terms, this weak scattering condition can be
understood as a condition for small quantum reflection from a
potential bump of linear size $\zeta$ and height $\av{V}$. It can be 
realized
in two ways: either the atomic kinetic energy $E$ is larger than
the correlation energy $E_\zeta$; then the potential
fluctuations must be small with respect to the atomic energy,
$\avV < E$. This case corresponds to the classical picture of
atoms flying well above small potential bumps. Or the atomic
energy is smaller than the correlation energy (requiring cooling
well below recoil, especially if $\alpha \ll 1$); then the
fluctuations must still be smaller than $E_\zeta$, but can
be larger than the atomic energy $E$. This case corresponds to a
quantum regime where the atom is able to tunnel through high
potential bumps of linear extension $\zeta$ thanks to its large de
Broglie wavelength $\lambda_\text{dB}=2\pi/k \gg \zeta$.

\paragraph{Scattering mean free path:}

\begin{figure}
\includegraphics[width=0.95\linewidth]{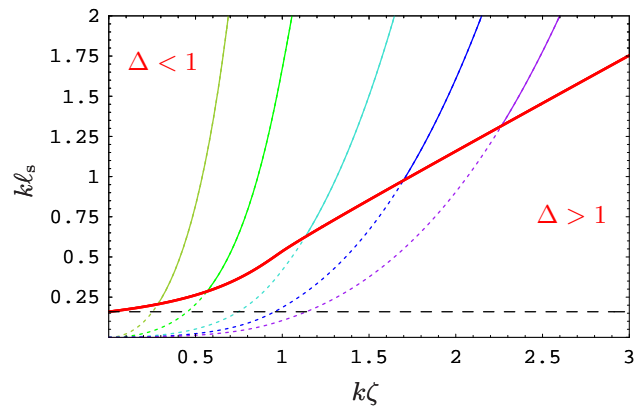}
\caption{(Color online) Plot of $k\ells$ from eq.~(\ref{ell_s.eq})
as a function of the reduced matter wave number
$k\zeta$ for different values of the disorder strength
$\eta=\avV/E_\zeta\in \{0.2,\,0.4,\,0.8,\,1.2,\,1.6\}$ (thin curves
from left to right). The thick line (red online), connecting points of
$k\ells$ where $\Delta=1$, i.\,e. $k\zeta=\sqrt{2}\,\eta$,
indicates the limit of validity of the weak scattering condition \ref{delta.eq}.
The dashed horizontal line corresponds to $k\ells=1/2\pi$.
A strongly disordered medium with $k\ells \approx 1$ is obtained for cold enough atoms.
}
\label{fig:smfp}
\end{figure}

In the weak scattering limit,
the average distance travelled by the atom between two scattering events defines the
elastic scattering mean free path $\ells$. It is
related to the speckle power spectrum through
\begin{equation}
\frac{1}{k\ells}=\Big(\frac{\eta}{k\zeta}\Big)^2 \int_{0}^{2\pi}
\frac{d\theta}{2\pi}\;\mathcal{P}(k\zeta,\theta), \label{ell_s.eq}
\end{equation}
where $\mathcal{P}(k\zeta,\theta)=8\calF(k\zeta\sin(\theta/2))$
represents the differential scattering cross-section. A plot of
$k\ells $ as a function of the reduced atomic wave number $k\zeta$
is shown in fig. \ref{fig:smfp} for different values of the
disorder strength $\eta$. The weak scattering condition $\Delta
\le 1$ implies the bound $k\ells \ge 1/2\pi$ such that the
shortest achievable scattering mean free path is of the order of
the 2D speckle correlation length $\zeta$ itself.
Fig.~\ref{fig:smfp} shows that the speckle potential, even though
correlated on the scale $\zeta$, can become a
highly disordered scattering medium with $k\ells$ of order
unity for sufficiently cold matter waves ($k\zeta \to 0$).

\paragraph{Boltzmann diffusion constant:}

While the atom propagates through the speckle field, it is
scattered by potential fluctuations. After a large number of
scattering events, this random walk conserving energy and particle
number results in diffusive transport for matter waves with wave
number $k$.
Following the approach pioneered by Vollhardt and W\"olfle
\cite{woelfle}, we set up a quantum kinetic equation
for the intensity kernel defined in
\refeq{Pqeps.eq} that allows to calculate the transport mean free
path from the microscopic properties of the system. In a first
step, we determine the 2D Boltzmann diffusion
constant
\begin{equation}
D_\text{B}(k)= \frac{\hbar k \,\elltr(k)}{2m},
\end{equation}
where the elastic
transport mean free path $\elltr$ is the
average distance travelled by the
atom before losing memory of its initial direction.
This approximation assumes that
the disorder average washes out all interference effects between
partial scattered waves. Scattering and
transport mean free paths are then
linked by the relation
\begin{equation}
\frac{\ells}{\elltr} = 1-\frac{\int_0^{2\pi} d\theta\,
\cos\theta\, \mathcal{P}(k\zeta,\theta)}{
\int_0^{2\pi}d\theta\,\mathcal{P}(k\zeta,\theta)}. \label{aniso}
\end{equation}
Because of the potential correlation at small scales
(Heaviside function in $\calF(x)$), the scattering
angle is bounded by $|\sin (\theta/2)| \le 1/k\zeta$. Hence,
there is no angular restriction for slow atoms
($k\zeta\le 1$). In the ultra-cold regime ($k\zeta \ll 1$) isotropic
scattering ($\elltr \approx \ells$) prevails, and the speckle potential
becomes an effective $\delta$-correlated potential. For
fast atoms $k\zeta\gg 1$, the maximum scattering angle is
$\theta_{\mathrm{max}} \simeq 2/k\zeta$ and scattering is strongly
peaked in the forward direction.
In this case, $\elltr \approx
(k\zeta)^2\,\ells \gg \ells$ (strongly anisotropic scattering).
For example, for Rubidium atoms with $k\zeta=1$, $L=2\,$cm,
$\alpha=0.1$, $P=0.25\,$W
and $\delta=10^6\,\Gamma$, we predict a diffusive matter wave
transport with elastic scattering mean free path $\ells \approx
2\,\mu$m and elastic transport mean free path $\elltr \approx
7\,\mu$m.

\paragraph{Weak localization:}

In phase coherent samples, the constructive interference between
counter-propagating amplitudes enhances the return probability of the
atomic matter wave to a given point.
This weak localization
correction \cite{altshuler,woelfle}
reduces the diffusion constant to $D=D_\text{B}-\delta D$ with
\begin{equation}
\frac{\delta D}{D_\text{B}} = \frac{2}{\pi} \frac{\ln
(L_0/\ells)}{k\elltr} \label{wl.eq}
\end{equation}
in 2D. Here, the length $L_0=\min(L,L_\phi)$ is the relevant
cutoff for fully coherent multiple scattering. It is either
the system size $L$ itself, or the phase coherence length $L_\phi
= \sqrt{D_\text{B}\tau_i}$ which accounts for possible phase breaking
mechanisms affecting interference at a rate
$\Gamma_i=\tau_i^{-1}$. For cold atoms in optical speckle
potentials, one phase breaking mechanism is inelastic scattering
of photons for which $\Gamma_i = \Gamma\,\av{V}/\hbar\delta$. The
corresponding phase coherence length $L_\phi$ scales as
$\delta^{2}I_\text{L}^{-3/2}$ while $\ells$ and $\elltr$ scale as
$\delta^{2}I_\text{L}^{-2}$. Thus, by keeping the ratio $I_L/\delta$ fixed,
all multiple scattering parameters $\av{V}$, $\ells$, $\elltr$ and
$D_\text{B}$ are kept constant. Under this condition, changing
$I_\text{L}$ (and $\delta$) only modifies $L_\phi$ and thus the interference
corrections. This opens the way to use inelastic scattering to monitor
the weak localization
corrections in a controlled manner,
just like with an external
magnetic field in 2D electronic experiments on negative
magnetoresistance \cite{Bergmann84}.

\begin{figure}
\includegraphics[width=.95\linewidth]{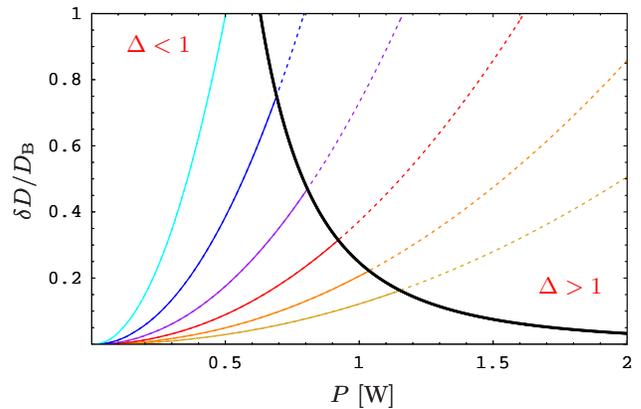}
\caption{(Color online) Weak localization corrections $\delta D$
relative to the Boltzmann diffusion constant $D_\text{B}$ at
detuning $\delta= 10^6\,\Gamma$, eq.~(\ref{wl.eq}), as a function of
laser power $P$ for different atomic matter wave numbers (from
left to right) $k\zeta=1.25,\,1.5,\,1.75,\,2.0,\,2.25,\,2.5$. The
speckle size is $L=2\,$cm, the aperture angle $\alpha=0.1$. For each value of $k\zeta$, the weak
scattering condition $\Delta<1$ is valid to the left of the thick
black line (solid curves).
Already for moderate laser power, weak localization corrections of
observable size are predicted.
\label{fig:ddsddp}}
\end{figure}

In fig.~\ref{fig:ddsddp}, we plot $\delta D/D_\mathrm{B}$
as a function of the total laser power $P$ for different values of
the atomic matter wave number $k$
at fixed laser
detuning $\delta=10^6\,\Gamma$. The thick black line indicates
the corresponding limit of validity of the weak scattering
condition $\Delta\leq 1$. The colder the atoms, the larger the
quantum corrections (within the lower bound $\zeta/L$ for $k\zeta$ and for
the applied laser power imposed by the diffusion condition $L_0 \ge \elltr$).
Even moderate laser
power assures \textit{coherent multiple scattering} in
the speckle plane and induces sizeable weak localization
corrections. For $k\zeta=2$, the relative correction $\delta
D/D_\text{B}$ attains approximately 30\% at $P = 0.9\,$W. For
$k\zeta=1.25$, the border $\delta D=D_\text{B}$ is even reached
within the region $\Delta \leq 1$ at $P = 0.5\,$W: this is the
strong localization onset.

\paragraph{Towards strong localization:}

\begin{figure}
\includegraphics[width=.95\linewidth]{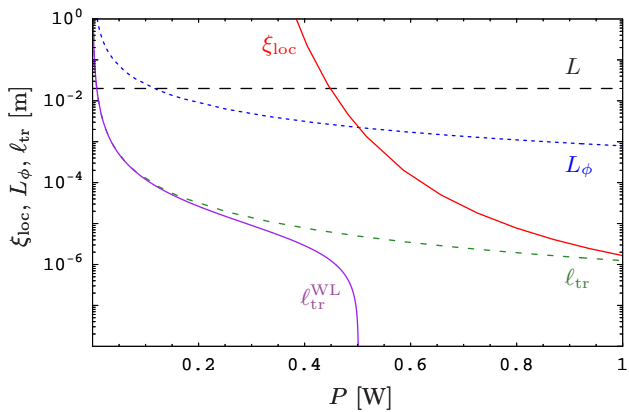}
\caption{(Color online) Logarithmic plot of the 2D localization
length $\xi_\text{loc}$ and of the phase coherence length
$\text{L}_\phi$ as a function of the laser power $P$ for
$k\zeta=1.25$ and $\delta=10^6\,\Gamma$.
The system size is fixed at $L=2\,$cm, the aperture angle at $\alpha=0.1$.
$\xi_\text{loc}$ and $L_\phi$ cross
at the strong localization threshold, which is reached for $P=
0.5\,$W, where the corrected transport mean free path
$\elltr^\text{WL}=2m(D_\text{B}-\delta D)/\hbar k$ vanishes.
\label{fig:lcmloc2d}}
\end{figure}

The perturbative weak localization correction (\ref{wl.eq}) diverges
with the cutoff length $L_0$, which
is compatible with the scaling prediction that waves are always
localized in 2D on sufficiently large coherent length scales
\cite{abrahams79}.
Equation (\ref{wl.eq}) is actually the result of a
self-consistent theory \cite{woelfle} that takes into account the relevant
singular terms driving the system towards the
strong localization onset $\delta D/D_\text{B} \to 1$.
The strong localization threshold is then reached when $L_0$ reaches the localization
length $\xi_{\text{loc}}\simeq \ells\,\exp(\pi k \elltr/2)$.

Because of its exponential growth, the
localization length becomes very large when $k \ells$
increases. Observing strong localization requires fully coherent
scattering in a sufficiently large speckle field. We therefore have to check
whether these requirements can be met with reasonable laser power
and atom temperatures. In fig. \ref{fig:lcmloc2d}, we have plotted
$\xi_{\text{loc}}$ and $L_\phi$ as a function of $P$ for
$k\zeta =1.25$, $L=2\,$cm, $\alpha=0.1$, and $\delta=10^6\,\Gamma$.
The two curves cross when
$P=0.5\,$W. At this point, we find for Rb$^{87}$ atoms
($\lambda_\text{L}=2\pi/k_\text{L}=0.78\,\mu$m)
$\xi_{\text{loc}}=2\,$mm, $\elltr=5\,\mu$m and
$\ell_s=0.92\,\mu$m. This places the strong localization threshold
at $k \ells \approx 0.93$ and $\eta \approx 0.77$, meaning that
the atoms have an energy slightly above the speckle fluctuations,
$E \approx 1.02 \, \avV$. The atomic temperature is then
$T=2.8\,$nK which is experimentally accessible
\cite{ketterle03}. The corresponding de Broglie wavelength is
$\lambda_\text{dB} = 8 \lambda_L \approx 6\,\mu$m.

\paragraph{To summarize:}

Making use of quantum transport theory, we have
studied the dynamics of ultracold atoms in 2D speckle potentials.
Starting from the microscopic potential correlation function, we
have calculated the elastic scattering mean free path and the
classical diffusion constant. Weak localization corrections are
shown to be of measurable size for realistic laser power and atom
temperature. We have found that the threshold to the strong
localization regime
is experimentally accessible and therefore worthwhile
further numerical and experimental investigation. We have given an estimate for the
localization length and provided a set of reasonable experimental
parameters that we hope will facilitate the realization in the
laboratory.

This work was supported by the DAAD, the DFG within SPP 1116,
the BFHZ-CCUFP and the Marie Curie program (contract number HPMT-2000-00102).
Laboratoire Kastler Brossel de l'Universit\'e Pierre et
Marie Curie et de l'\'Ecole Normale Sup\'erieure is UMR 8552 du
CNRS.

\end{document}